\documentclass[sn-aps,Numbered,pagebackref=false]{sn-jnl}
\usepackage[T1]{fontenc}
\usepackage{geometry}
\usepackage[ngerman,english]{babel}
\usepackage{amsmath}
\usepackage{amssymb}
\usepackage{empheq,fancybox}

\usepackage{bibspacing}
 \usepackage{hyperref}

\newcommand{\eref}[1]{(\ref{#1})}

\newcommand{\Liou}{\mathcal{L}}
\newcommand{\mcP}{\mathcal{P}}
\newcommand{\mcQ}{\mathcal{Q}}
\newcommand{\mcF}{\mathcal{F}}
\newcommand{\mcM}{\mathcal{M}}

\newcommand{\WP}{W_\mcP}
\newcommand{\WQ}{W_\mcQ}
\newcommand{\GP}{G_\mcP}
\newcommand{\FP}{F_\mcP}

\newcommand{\tr}{{\mathrm{tr}}}
\newcommand{\otr}{\overline{\tr}}

\begin{document}
\setlength{\baselineskip}{3.75mm}

\title{Entropy augmentation through subadditive excess:\\
information theory in irreversible processes}

\author*[1]{\fnm{J\"{u}rgen} \sur{Stockburger}}\email{juergen.stockburger@uni-ulm.de}

\affil[1]{\orgdiv{Institute for Complex Quantum Systems and IQST}, \orgname{Ulm University}, \orgaddress{\street{Albert-Einstein-Allee 11}, \postcode{89081} \city{Ulm}, \country{Germany}}}

\abstract{Within its range of applicability, the Boltzmann equation seems unique in its capacity to accurately describe the transition from almost any initial state to a self-equilibrated thermal state. Using information-theoretic methods to rephrase the key idea of Maxwell and Boltzmann, the \emph{Sto\ss{}zahlansatz}, a far more general, abstract ansatz is developed. An increase of the Gibbs-Shannon-von Neumann entropy results without the usual coarse-graining. The mathematical structure of the ansatz also provides avenues for efficient computation and simulation.}

%\keywords{Irreversibility, Boltzmann equation, information entropy, equilibration}

\maketitle

\section{Introduction}

Among the many approaches relating irreversible physical phenomena to dynamical laws at  microscopic level, the Boltzmann equation~\cite{boltz72,cerci88} still stands out for both its simplicity and its wide applicability. Within its clear limits of validity -- diluteness of a gas and short-range scattering interactions -- it seems almost unassailable, as far as experimental tests go.
It has proven itself as a successful modeling tool in applied fields as diverse as the atmospheric reentry of spacecraft~\cite{baran17} and transport of charge carriers and other out-of-equilibrium excitations in semiconductors~\cite{snoke11}. States very far from equilibrium play a key role in both cases, and they appear to be handled very well in the clear and simple physical picture drawn by Boltzmann and Maxwell. In some applications, versions extended to quantum gases~\cite{uehli33,green53} are used.

Almost immediately, the manifestly irreversible character of the Boltzmann equation led to epistemological questions regarding the potential identification of micro- and macrostates as well as the role of chance in this connection. The necessity -- and the difficulty -- of reconciling the manifest irreversibility of processes in the macroscopic world with microreversibility became clear, and a scientific and philosophical debate spanning centuries ensued.

Long before formal information theory, the hypothetical entity known as Maxwell's demon was invented~\cite[p. 308]{maxwe71}, with the precise purpose of questioning the view of the Second Law as an axiom. On the other hand, Maxwell sees our manifest \emph{in}ability to perform the task assigned to the demon as crucial for the emergence of thermodynamics. Later work by Boltzmann~\cite{boltz77}, reviewed by Ehrenfest and Ehrenfest~\cite{ehren11}, takes incomplete information into account by associating with any point in phase space a surrounding domain as a macrostate, whose volume determines its thermodynamic entropy. In this later picture, dynamics is expressed only in terms of pure state, and an increase of entropy results if the phase space volume of the macrostate associated with the current phase space point expands over time. This seems difficult to reconcile with the Boltzmann equation, where changes in the irreversibility measure ``$H$'' are an intrinsic feature of dynamics in terms of a mixed state. The differences and commonalities between various approaches to statistical physics has been reviewed by Goldstein et al.~\cite{golds20}. The Maxwellian perspective that incomplete information is the key to the emergence of irreversibility is not restricted to the specifics of the Boltzmannian approach to statistical mechanics.

With explicit reference to Maxwell, Myrvold~\cite{myrvo11} develops a view of thermodynamic entropy as \emph{means-relative}, which seems appropriate since the thermodynamic state inferred from observations should depend on the level of detail provided by \emph{thermodynamic observations}. Long before formal information theory, Smoluchowski~\cite{smolu04} discusses this idea at length in his contribution to the Festschrift dedicated to Boltzmann on the occasion of his sixtieth birthday~\cite{meyer04}. A more recent work with at least hints in this direction is \cite{golds20}.

It is instructive to re-formulate this perspective as \emph{means-relative irreversibility}:
The progress of real (although approximate) experimental reversals of time, ranging from the famed Hahn echo~\cite{hahn50} to a recent paper on time-reversal in qubits~\cite{schia23}, seems motivation enough to seriously consider the view that thermodynamic entropy may be means-relative. The Boltzmann equation seems to be quite compatible with this perspective, since its key element, \emph{Sto\ss{}zahlansatz}, is an approximation very unlikely to distort any thermodynamic observations at later times, whatever the means may be~\footnote{Of course, this statement is made with the proviso that the Boltzmann equation is applied within its domain of validity, i.e., on a dilute gas with short-range interactions.}. The idea of interpreting irreversibility as a loss if information~\cite[p.~72]{born49}, particularly information on correlations~\cite{prigo70}, had prominent adherents through the years. This concept is also implicitly realized in the Boltzmann equation and its extension into the BBGKY hierarchy~\cite{cerci97,campa14}, but is revealed to be far more general when viewed through the lens of information theory. More recent experiments on spin echos~\cite{levst98,pasta00} and their relevance in the study of information loss and decoherence~\cite{zanga17} indicate that a more general approach to irreversibility is needed than scattering-based theories.

While the link between information theory and statistical physics  has been thoroughly explored in the context of equilibrium thermodynamics~\cite{jayne57}, information theoretic methods appear to have been used more sparingly in the context of irreversible dynamics. However, the maximization of entropy under suitable constraints or for a subsystem is a defining feature for some approaches~\cite{lewis67,grabe78,oettin10}. The microstate resulting from such a maximization strongly depends on any imposed constraints, i.e., on the set of observables elevated to the status `thermodynamic observable'~\cite{smolu04,lewis67}. Most of these approaches share the assumption that irreversibility can be introduced by rapidly relaxing a ``fast'' subsystem exempt from reversible microdynamics.

In current research, an opposite view is often pursued: The paradigm of canonical typicality~\cite{golds06,popes06,gemme09} posits that the emergence of thermal properties of a subsystem is a generic feature of entanglement in large systems. Similarly, and echoing Boltzmann's later view~\cite{boltz77}, the eigenstate thermalization hypothesis~\cite{deuts18} keeps microscopic dynamics untouched, seeking an explanation of thermodynamic phenomena as arising from particular modes of preparation and observation. A related work of von Neumann\cite{neuma29}, often cited as an ancestral paper for this line of inquiry, makes the link to Boltzmann's work explicit.

%+++ONCE VS. INTERLEAVED+++

The Boltzmann equation is not associated with a concrete set of thermodynamic observables. Since it propagates the single-particle phase-space density, \emph{all} conceivable single-particle observables can be obtained from its solution. The formal entropy increase implied by Boltzmann's H-Theorem is therefore not necessarily identical to an increase in thermodynamic entropy at each instant. However, it clearly indicates the irreversible nature of time propagation under the Boltzmann equation. In this work, the negative of Boltzmann's H is identified as the information entropy of a marginal distribution. This results in a particularly concise proof of the H-Theorem, which rests entirely on the observation that the Boltzmann equation eliminates correlations through the implicit assumption that particles can be treated as independently distributed at any time. The key step of the proof rests on the fact that mutual information is non-negative, a fact that is also known as subadditivity of entropy in the context of physics.

Eliminating correlations thus leads to a subadditive excess of entropy, augmenting it in infinitesimal steps over time. This can be formulated as an abstract principle, the EASE principle (Entropy Augmentation through Subadditive Excess). EASE, in essence, preempts the operational impossibility to distinguish the correlated from the decorrelated state after some time has passed. Depending on personal taste, this postulated impossibility could be seen as a somewhat more  abstract and general reformulation of the molecular chaos hypothesis. In the quantum regime, it can be seen as a general, system-agnostic approach aiming in a similar direction as studies of entanglement dynamics~\cite{ho17,chan19} and scrambling of quantum information~\cite{sekin08,mi21} in quantum many-body physics.

The paper is organized as follows: Section 2 gives an overview of the information-theoretic prerequisites of this work and applies them to the Boltzmann equation, the \emph{Sto\ss{}zahlansatz} and the H-Theorem. Section 3 translates the key ideas of EASE into equations of motion of different forms, notably a non-linear generalization of the Nakajima-Zwanzig equation, suitable for analytic approximation, and a version involving a system of linear first-order equations, suitable for numerical evaluation. Sections 4 and 5 provide a discussion of results and concluding remarks.

\section{Information theory in the context of dynamics}

\subsection{Information-theoretic prerequisites}

The basic definition of entropy in this work is that of Gibbs, Shannon and von Neumann. For consistency with thermodynamics, the natural logarithm is used, and Boltzmann's constant appears as a prefactor. We have
\begin{equation}
  S[{p_i}] = - k \sum_i p_i \log p_i = - k \mathbb{E}\left[ \log p_i \right]
\end{equation}
for a discrete classical probability space ($\mathbb{E}$ denotes expectation values) and the expression
\begin{equation}
  S[W] = - k \tr W log W
\end{equation}
for a quantum system described by density matrix $W$~\footnote{For continuous classical distributions, see the introduction of relative entropy below.}. $S$ quantifies the average information to be gained from the outcome of a measurement when the distribution or density matrix is known beforehand.

Maximizing this entropy under constraints set by conservation laws and/or measurements of observables yields mixed states which can be identified with thermodynamic states -- a view which can already be found in the work of Boltzmann~\cite{boltz77}, long before formal information theory.
% see klein61, p. 14
For equilibrium states, a clear link between information entropy and thermodynamic entropy has been demonstrated by Jaynes~\cite{jayne57}.

The potential discrepancy between information entropy and thermodynamic entropy in a \emph{dynamical} setting can be illustrated by a simple information-theoretic
analysis of pseudo random number generators: These algorithms are purely deterministic, i.e., their output has zero information entropy~\footnote{If its seed value is random, it has finite entropy, but far less than the entropy limited tests would ascribe to the output.}, but a barrage of empirical tests would indicate that the output is consistent with a stream of random, independent, unbiased binary digits, a high-entropy output signal.

A similar discrepancy exists in physical dynamical systems, and the analogy between random number generators and dynamical systems has actually prompted the development of random number generators which mimic dynamical chaos as closely as possible~\cite{james20} in order to make patterns of correlations virtually undetectable; empirical tests of the output would ascribe it a high entropy. This seems analogous to a thermodynamic entropy ascribed to a complex dynamical system after time propagation. In this case the physical entropy is typically higher than the Gibbs-Shannon-von Neumann entropy after reversible propagation.

The growth of entropy through thermodynamic state changes needs to be reconciled with the microscopic dynamics. One must either abstain from the identification of information entropy with thermodynamic entropy in the case of irreversible processes or conclude that microscopic dynamics alone are an insufficient description of thermodynamic state changes. In the present work, a compromise between these positions will be sought, the logical rationale being that the conservation of Gibbs-Shannon entropy can also serve as justification for ``backdating'' all or part of the information loss implied in some later transition to the thermodynamic picture.

The microscopic laws make predictions which are simply too precise (in the sense of information theory) to be complete, realistic description of the information transmitted in the complete sequence ``preparation $\to$ propagation $\to$ thermodynamic measurement''. where the last step discards the majority of microscopic information. 
We see, at best, the marginal distributions of a handful of observables. For intermediate states, there is a mismatch between the dynamical state \emph{pre}dicted from preparation and deterministic dynamics on the one hand, and \emph{retro}diction from the observables accessible at a later time.
A pragmatic stance is taken here: The question pursued is not where and how information is lost, but at what stage some information can be ``deleted'' because it will not be retrievable at any later time.

 When irreversible processes are viewed as similar to mixing process~\footnote{Planck writes in 1933 \cite[p. 274]{planc33}:
  \foreignlanguage{ngerman}{``Nach der atomistischen Idee gleicht der W\"arme\"ubergang von einem w\"armeren zu einen k\"alteren K\"orper nicht dem Herabsinken eines Gewichtes, sondern etwa einem Mischungsvorgang, der darin besteht, da\ss{} zwei verschiedene, in einem Gef\"a\ss{} befindliche Sorten Pulver, die anfangs \"ubereinander geschichtet sind, beim fortgesetzten Sch\"utteln des Gef\"a\ss{}es  sich allm\"ahlich vermengen.''}; English translation~\cite[p. 96]{planc36}:
``According to the atomist idea the transference of heat from the hotter to the colder body does not resemble the falling of a weight; what it resembles is a mixing process, as when two different kinds of powder in a vessel, having first constituted different layers, eventually mingle with each other if the vessel is continually shaken''}, such backdating seems more natural than maintaining reversibility until a sudden transition to a thermodynamical perspective happens at some measurement event. We will see that the Boltzmann equation performs this task of backdating in an efficient and elegant manner, with information loss spread out continuously over time. Pointing out ways to generalize this feature is the main objective of this work.

If the probability space or Liouville space can be decomposed a product space (assuming $n$ factors), additional formal definitions relate the entropy $S$ with the entropies $S_j$ of each part. The \emph{mutual information} of the parts, $I_{1,\ldots,n}$ denotes an excess due to the subadditivity of entropy, being defined through
\begin{equation}
S = S_1 + \ldots + S_n - I_{1,\ldots,n}.
\end{equation}
When mutual information cannot be inferred from measurements or observations, a system appears to have the entropy $S_1 + \ldots + S_n$ instead of $S$, typically being larger by a \emph{subadditive excess}~\footnote{One might find this terminology redundant, but it is introduced to stress the link between neglected mutual information and increase of entropy}.
Formally, mutual information is a Kullback-Leibler divergence of two distributions, defined as
\begin{align}
  D(p\Vert q) &=  k \sum_i p_i \log(p_i/q_i) &\textrm{(classical)}\\
  D(W\Vert V) &=  k \,\tr W \left(\log W - \log V \right) &\textrm{(quantum)}.
\end{align}
Mutual information is the special case where the second argument is a product of the marginal distributions (or reduced density matrices) of the $n$ parts, derived from the first argument,
\begin{align}
  I_{1\ldots n}  &= D(p\Vert p^{(1)}\cdot p^{(2)}\cdots p^{(n)}) &\textrm{(classical)}\\
  I_{1\ldots n}  &= D(W\Vert \rho_1\otimes \rho_2\otimes\cdots\otimes \rho_n) &\textrm{(quantum)}.
\end{align}
The Kullback-Leibler divergence is non-negative, and it is zero if and only if both arguments are the same distribution. As a consequence, mutual information is non-negative, and it is zero only if all parts it describes are mutually independent.
Moreover, the Kullback-Leibler divergence has a well-defined continuum limit. Since the second argument need not be normalized, it can be used to define information entropy as relative entropy in a continuous probability space (relative to a background measure, e.g., the Lebegue measure).

For quantum dynamics under the influence of (classical) random forces, the above definitions can be applied to the density matrix of the resulting mixed state. At a finer level of granularity, considering the history of the random force $\xi_t$ as a classical observable, a joint entropy of the noise source and the noise-driven state is given by
\begin{equation}\label{eq:jointQnoise}
  S(W,\xi) = - k \mathbb{E}\left[ \tr W[\xi_t] \log(p\left[\xi_t] W[\xi_t] \right)\right],
\end{equation}
% W[\xi_t] \right)\right],
where $ W[\xi_t]$ is the conditional state resulting from dynamics under the noise realization $\xi_t$. $S(W,\xi)$ is a mixed quantum-classical form of joint entropy. Subtracting from this the entropy of the noise source, $S[\xi_t] = - k \mathbb{E}[\log p[\xi_t]]$, we obtain the \emph{conditional entropy}
\begin{equation}\label{eq:condQnoise}
  S(W|\xi) = S(W,\xi) - S(\xi)
  = - k \mathbb{E}\left[\tr W[\xi_t] \log W[\xi_t] \right],
\end{equation}
the globally averaged logarithm of the conditional density matrix. Since the conditional state $ W[\xi_t]$ is simpler to analyze than its expectation value, it can be a useful quantity.
%For random unitary propagation of a pure state, e.g., we have $S(W|\xi)\equiv 0$.

Mutual information $I(W;\xi)$ between quantum system and noise source can be defined through
\begin{equation}\label{eq:mutualQnoise}
S(W,\xi) = S(W) + S(\xi) - I(W;\xi),
\end{equation}
where the marginalization of $W$ in $S(W)$ is defined through $W = \mathbb{E}\left[ W[\xi_t] \right]$. Any noise-averaged quantum dynamics thus leads to a density matrix with quantum entropy
\begin{equation}
S(W) = S(W|\xi) + I(W;\xi) = I(W;\xi)\label{eq:condplusmutual}
\end{equation}
which is identical to the mutual information between system state and noise: The first equality results from combining Eqs. (\ref{eq:condQnoise}) and (\ref{eq:mutualQnoise}), while $S(W|\xi)\equiv 0$ follows from the fact that the propagation is unitary for any fixed noise realization.

$I(W;\xi)$ is additive for independent noise contributions taking effect consecutively; so $I(W;\xi)$ is a non-decreasing function of time, at least for a Markov process; the same holds for $S(W)$, of course. This may be one way a nominally isolated system equilibrates toward thermal equilibrium, if the rate at which the noise coupling augments the entropy of the quantum system far exceeds the rate at which it transfers energy.

Without an external noise source, there is no obvious ``information sink'' responsible for the increasing entropy in an irreversible process. As is done in virtually all formal descriptions of irreversibility, one can make use of the non-uniqueness of a microstate associated with a given macrostate or a given set of observables. Naively choosing entropy maximization at every instant as the criterion defining the connection between \emph{dynamical} micro- and macrostates can be successful, but the results may be highly specific to the set of macro-observables chosen~\cite{lewis67,grabe78,oettin10}.

The Boltzmann equation cannot be derived strictly in this way: Here it is crucial that entropy augmentation is performed after, not during scattering events~\footnote{Continuous application of entropy maximization, constrained by the complete set of single-particle observables, leads to mean-field dynamics (Vlasov equation)~\cite{lewis67}}. Correlations are allowed as \emph{transient} features of the distribution which are ``forgotten'' between collisions. In some sense, \emph{internal} correlations may perform the function of an information sink from which information cannot be recovered.

\subsection{Information-theoretic contextualization of Boltzmann's H-theorem}
Boltzmann's transport equation for dilute gases has the form of a single-particle Liouville equation with an additional \emph{nonlinear} term representing scattering events whose duration is considered negligible compared to other timescales, the single-particle phase-space density $\rho(\mathbf{r} ,\mathbf{p},t)$ obeys
\begin{equation}
\frac{\partial \rho}{\partial t}+\frac{\mathbf {p}}{m}\cdot \nabla \rho+\mathbf{F} \cdot {\frac{\partial \rho}{\partial \mathbf{p} }}
=\left(\frac{\partial \rho}{\partial t}\right)_{\mathrm {coll}}.
\end{equation}
For elastic scattering, the collision term reads
\begin{align}
\left({\frac {\partial \rho}{\partial t}}\right)_{\text{coll}}
=\iint &gI(g,\Omega )[\rho(\mathbf{r} ,\mathbf{p'},t)\rho(\mathbf{r} ,\mathbf {p'}_{B},t) %\nonumber\\
-\rho(\mathbf{r} ,\mathbf{p},t)\rho(\mathbf{r} ,\mathbf{p}_{B},t)]\,d\Omega \,d^3 p_{B},
\end{align}
where the momenta $\mathbf{p}$ and $\mathbf{p}_{B}$ ($\mathbf{p'}$ and $\mathbf{p'}_{B}$) refer to two particles before (after) scattering. These momenta are linked by momentum and energy conservation: The momenta after scattering are completely determined by $\mathbf{p}$ and $\mathbf{p}_{B}$, the absolute value $g = |\mathbf{p}_B - \mathbf{p}| = |\mathbf{p'}_B - \mathbf{p'}|$ of the momentum difference, and a point $\Omega$ on the unit sphere representing the scattering angle. $I(g,\Omega)$ denoted the differential cross section of the scattering problem.

The Boltzmann equation is not an exact description of microscopic dynamics. It contains the famed \emph{Sto\ss{}zahlansatz}, signified by the appearance of the single-particle density in the form of a product with itself, rather than a many-particle density reflecting the correlations built up over previous collision events.

This approximation is needed to obtain a closed equation, and it is instrumental in giving this equation its characteristics of irreversibility and equilibration. Irreversibility is expressed by the statement that the single-particle entropy
\begin{equation}
S_1(t) = -\iint \rho(\mathbf{r} ,\mathbf{p},t) \log \rho(\mathbf{r} ,\mathbf{p},t) d^3r d^3 p,
\end{equation}
is non-increasing, equivalent to Boltzmann's H-Theorem. Boltzmann and many subsequent authors relied heavily on physical arguments, scattering theory in particular, when proving the theorem.

In the following, a concise information-theoretic proof of the H-theorem will be given. Consider an initially uncorrelated phase-space density $W(\{q_j\},\{p_j\},t) = \prod_j \rho(q_j,p_j,t)$ and let it evolve for some time $\Delta t$, which is smaller than the collision-free time yet larger than the interaction time of a scattering event. At the end of the time step, the Gibbs-Shannon entropy will be unchanged, but $W(\{q_j\},\{p_j\},t+\Delta t)$ will (in most cases) no longer describe independent particles. \emph{Modifying} the dynamics by deliberately making the substitution $W(\{q_j\},\{p_j\},t+\Delta t)$ $\to$ $\prod_j \rho(q_j,p_j,t+\Delta t)$ will \emph{increase} the Gibbs-Shannon entropy $S$ of the phase-space density unless correlations are absent at time $t+\Delta t$, e.g., in a fully thermalized state. This is a consequence of the subadditivity of entropy, or, equivalently, the non-negativity of mutual information. Since $S$ is non-decreasing, $H = -S/N$ is non-increasing. It remains to be shown that this construction in terms of finite time steps yields the Boltzmann equation in the usual limits. The choice of time step $\Delta t$ made here allows scattering theory to be applied, thus continuous propagation of the Boltzmann equation over the interval $[t,t+\Delta t]$ should differ only by a term $o(\Delta t)$ from the result for $\rho(q,p)$ resulting from the stepwise Liouville-propagation-then-decorrelate procedure outlined here. Then applying the limit $\Delta t\to 0$ to the stepwise propagation over a larger interval divided in steps of size $\Delta t$ converges to a scenario where $W(\{q_j\},\{p_j\},t) = \prod_j \rho(q_j,p_j,t)$ at all times, and where $\rho(q,p,t)$ obeys the Boltzmann equation.

It is now a key observation that details of scattering theory are not essential for non-decreasing entropy, but the \emph{Sto\ss{}zahlansatz} is, \emph{merely for the fact that it discards correlations}. This raises the question whether it should be considered a representative of a wider class of `decorrelating approximations' which are suitable to represent our incomplete knowledge of an imperfectly controlled and observed complex system.

\section{Entropy Augmentation through Subadditive Excess}

\subsection{Decorrelating projection operator}
In the example of the Boltzmann equation we have seen that decorrelating the many-particle density adds a ``subadditive excess'' to the entropy. This will be developed into a formal, far more general concept, which will be called
Entropy Augmentation through Subadditive Excess (EASE principle).

Similar to the decomposition of a many-particle phase space into $N$ single-particle terms, we will consider a more abstract division of a Liouville space into $n$ partitions, whose product space is the original Liouville space. What these partitions might be is left open here. The number of partitions may be as small as $n=2$, or as large as in the Boltzmann equation.
For the sake of argument, we will work with a quantum Liouville equation, but the classical case is closely related.

We start from the linear dynamics
\begin{equation}
\dot{W} = \Liou W
\end{equation}
where $W$ is the density operator of the composite system and $\Liou$ its Liouville operator. Its $n$ partitions have reduced density operator $\rho_j$, obtained by a partial trace over all partitions other then $j$, for which we introduce the notation
\begin{equation}
  \rho_j = \otr_jW. \label{eq:otr}
\end{equation}
We now define the decorrelating projector $\mcP$ as the map
\begin{equation}
\mcP: W \mapsto \rho_1\otimes\rho_2\otimes \cdots \otimes\rho_n .
\end{equation}
The term `projector' seems justified since $\mcP(\mcP W) = \mcP W$. For any correlated state, application of $\mcP$ raises its entropy. The projector $\mcP$ is \emph{nonlinear}~\footnote{A quasilinear expression for $\mcP$, parameterized by the reduced density matrix, was introduced for the case $n=2$ in~\cite{willi74}.} since it is a product of the $\rho_j$, each of which depends linearly on $W$. This is a feature which complicates the application of the standard Nakajima-Zwanzig  projection formalism~\cite{nakaj58,zwanz60}.

The utility of introducing the projector $\mcP$ into the microscopic dynamics will therefore be first demonstrated in a transparently constructed elementary example: Let us consider a dynamics where on each infinitesimal time step $dt$ either the Liouville equation is followed with probability $1-\gamma dt$ or the state is projected to  $\mcP W$ with probability $\gamma dt$. The dynamics is thus described by a stochastic differential equation driven by a Poisson process $N_t$ with rate $\gamma$,
\begin{equation}
  d W_N = \Liou W_N dt + (\mcP W_N - W_N) d N_t. \label{eq:stochdN}
\end{equation}
On any infinitesimal interval $[t,t+dt]$, the differential $d N_t$ denotes a random variable which equals $1$ with small probability $\gamma dt$ and equals $0$ with probability $1-\gamma dt$.

In less formal terms, the dynamics described by \eref{eq:stochdN} follows the standard Liouville equation for intervals for some random duration, with ``jumps'' $W_N \to \mcP W_N$ occurring between the intervals. The waiting times $\tau$ between jumps are independent and exponentially distributed, $p(\tau) = \gamma \exp(-\gamma\tau)$%~\footnote{Random waiting times avoid potential artifacts caused by a fixed sampling frequency.}
.

For any fixed realization $N_t$, the entropy associated with the density matrix $W_N$ increases, except in the case of no jumps whatsoever (with small probability $\exp(-\gamma t)$), or when $\mcP W_N = W_N$ for every jump. Thus we have a non-decreasing conditional entropy $S(W|N_t) = - k\mathbb{E}[\tr W_N \log W_N]$. With
\begin{equation}
  S(W) = S(W|N_t) + I(W;N_t),
\end{equation}
analogous to the first equality in Equation \eref{eq:condplusmutual}, the Markov property of the Poisson process and the non-negativity of $I(W;N_t)$, one concludes that $S(W)$ is bounded from below~\footnote{This is actually a somewhat weaker property than $S(W)$ itself non-decreasing, but the difference is of little concern here.}
by the non-decreasing function $I(W;N_t)$.

As an alternative to Equation \eref{eq:stochdN}, let us next consider the deterministic equation
\begin{equation}
\frac{dW}{dt} = \Liou W - \gamma\cdot(W - \mcP W ) \label{eq:meandN}
\end{equation}
where $W$ can be identified with the expectation value of $W_N$. Equation \eref{eq:meandN} can be seen as the mean-field approximation of Equation~\eref{eq:stochdN} with respect to the noise statistics~\footnote{Since $\mcP$ is nonlinear, the operations $\mathbb{E}$ and $\mcP$ cannot be interchanged.}. Evidently, the Liouville equation is recovered for $\gamma\to 0$. Moreover, one easily demonstrates that the entropy associated with a density matrix $W$ governed by \eref{eq:meandN} is non-decreasing. Since the microscopic Liouvillian conserves information entropy, the entropy change associated with Equation \eref{eq:meandN} is
\begin{align}
  \dot{S} & %= - \tr\dot{W} \log W \nonumber\\
  = \gamma\, \tr W \log W - \gamma\, \tr \WP \log W\nonumber\\
  &= \gamma\, \tr W \log W - \gamma\, \tr \WP \log \WP %\nonumber\\
  + \gamma\, \tr \WP \log \WP - \gamma\, \tr \WP \log W\nonumber\\
  &= \gamma\, \tr W \log W - \gamma\, \tr W \log \WP %\nonumber\\
  + \gamma\, \tr \WP \log \WP - \gamma\, \tr \WP \log W\nonumber\\
   &= \gamma\, D(W\Vert\WP) + \gamma\, D(\WP\Vert W) \geq 0,
\end{align}
where the change from the second to the third line makes use of the fact that the logarithm of a factorizing density matrix has the algebraic structure of a sum of single-particle observables, i.e., a trace with weight $\WP$ is the same as a trace with weight $W$.

\subsection{Projection formalism: relevant and residual density matrix}

In the spirit of the usual derivation of the Nakajima-Zwanzig equation~\cite[sec.~9.1]{breue02}, an exact system of coupled equations of motion for a relevant density matrix $\WP$ and a residual density matrix $\WQ$ can be found%~\footnote{In most of the literature, $\WQ$ is called \emph{irrelevant}. In this work, this designation does not apply literally, since $\WQ$ will not always be neglected.}
. We start from the Liouville equation and the definitions
\begin{equation}
  \WP = \mcP W\quad\mbox{and}\quad \WQ = W - \WP,      \label{eq:WPWQ}
\end{equation}
aiming to find coupled equations of motion for $\WP$ and $\WQ$.
A dynamical equation for $\WP$ is obtained through the product rule
\begin{equation}
  \dot\WP = \dot\rho_1\otimes\rho_2\otimes\cdots\otimes\rho_n
  + \rho_1\otimes\dot\rho_2\otimes\cdots\otimes\rho_n
  + \cdots
  + \rho_1\otimes\rho_2\otimes\cdots\dot\rho_n.
\end{equation}
Using Liouville's equation, this can be restated in the form
\begin{align}
  \dot\WP &= \otr_1\Liou W\otimes\rho_2\otimes\cdots\otimes\rho_n
  + \rho_1\otimes\otr_2\Liou W\otimes\cdots\otimes\rho_n
  + \cdots %\nonumber\\
  + \rho_1\otimes\rho_2\otimes\cdots\otr_n\Liou W
\end{align}
or
\begin{align}
  \dot\WP &= \otr_1\Liou\WP\otimes\rho_2\otimes\cdots\otimes\rho_n
  + \rho_1\otimes\otr_2\Liou\WP\otimes\cdots\otimes\rho_n
  + \cdots %\nonumber\\
  + \rho_1\otimes\rho_2\otimes\cdots\otr_n\Liou\WP\nonumber\\
  & + \otr_1\Liou\WQ\otimes\rho_2\otimes\cdots\otimes\rho_n
  + \rho_1\otimes\otr_2\Liou\WQ\otimes\cdots\otimes\rho_n
  + \cdots %\nonumber\\
  + \rho_1\otimes\rho_2\otimes\cdots\otr_n\Liou\WQ.
\end{align}
The formal dynamics of $\WQ$ is essentially obtained by subtracting this equation from Liouville's equation with $W = \WP + \WQ$ inserted,
\begin{align}
  \dot\WQ &= \Liou\WQ
  - \otr_1\Liou\WQ\otimes\rho_2\otimes\cdots\otimes\rho_n %\nonumber\\
 - \rho_1\otimes\otr_2\Liou\WQ\otimes\cdots\otimes\rho_n
   - \ldots
  - \rho_1\otimes\rho_2\otimes\cdots\otr_n\Liou\WQ\nonumber\\
& + \Liou\WP
  - \otr_1\Liou\WP\otimes\rho_2\otimes\cdots\otimes\rho_n %\nonumber\\
 - \rho_1\otimes\otr_2\Liou\WP\otimes\cdots\otimes\rho_n
   - \ldots
  - \rho_1\otimes\rho_2\otimes\cdots\otr_n\Liou\WP.
\end{align}
These are closed equations of motion for the pair $(\WP,\WQ)$ since $\rho_j$ is a function of $\WP$ as a consequence of Equation \eref{eq:otr} and the projection property of $\mcP$.

With the definition
\begin{align}
  \mcF(U,V)
& = \otr_1\Liou V\otimes\otr_2 U\otimes\cdots\otr_n U\nonumber\\
& + \otr_1 U\otimes\otr_2\Liou V\otimes\cdots\otr_n U
  + \cdots\nonumber\\
& + \otr_1 U\otimes\otr_2 U\otimes\cdots\otr_n\Liou V
\end{align}
and
$%\begin{equation}
  \mcM(U) = \mcF(U,U)
$%\end{equation}
the set of dynamical equations attains the compact form
\begin{align}
  \dot\WP &= \mcM(\WP) + \mcF(\WP,\WQ)\label{eq:eomP}\\
  \dot\WQ &= \Liou\WQ - \mcF(\WP,\WQ) + \Liou\WP - \mcM(\WP).\label{eq:eomQ}
\end{align}
These equations are nonlinear in $\WP$, but Equation~\eref{eq:eomQ} is linear in $\WQ$. Therefore its solution can be represented as
\begin{align}
  \WQ(t) &=   \int_0^t ds \GP(t,s) \left( \Liou \WP(s) - \mcM(\WP(s)) \right)
%\nonumber\\
+ \GP(t,0) \WQ(0)
  \label{eq:WQsolution}
\end{align}
where the kernel $\GP$ is defined as the solution of the initial value problem
\begin{align}
  \partial_t \GP(t,t') &= (\Liou-\FP) \GP(t,t');\quad
  \GP(t',t') = \operatorname{id}
\end{align}
and where $\FP$ is the linear map $V \mapsto \mcF(\WP,V)$. After using Equation \eref{eq:WQsolution} to eliminate $\WQ$, the nonlinear Nakajima-Zwanzig equation
\begin{align}
  \dot\WP &= \mcM(\WP) + \FP
  \int_0^t ds \GP(t,s) \left( \Liou \WP(s) - \mcM(\WP(s)) \right)\nonumber\\
  &+ \FP \,\GP(t,0) \WQ(0)
  \label{eq:NZnonlin}
\end{align}
for the relevant part $\WP$ results. Equation \eref{eq:NZnonlin} is one of the central results of this paper.

While the notation of Equation \eref{eq:NZnonlin} is highly reminiscent of the known linear case, it must be stressed that Equation \eref{eq:NZnonlin} is much more complex since $\GP(t,s)$ is a nonlinear functional of the relevant state $\WP$ on the entire interval $[s,t]$. But the example of the Boltzmann equation shows that physically meaningful approximations of the memory term are possible.

\subsection{Finite correlation memory}

A generic strategy in the spirit of \emph{Sto\ss{}zahlansatz} but far more general is the introduction of a cutoff function \emph{limiting the memory time} $t-s$, a highly plausible simplification in the context of many-particle systems.
When introducing such approximations outside of a concrete physical context, it appears easier to discuss them as modifications to the equation of motion \eref{eq:eomQ} of the residual density matrix instead of directly modifying $\GP$.

In the context of approximations on $\WQ$, the role of $\WP$ and $\WQ$ as state variables of a dynamical system will be given precedence over Equation \eref{eq:WPWQ}, which will apply only to initial values in this case: After approximations on the dynamical laws, $\mcP W(t)$ and $\WP(t)$ are not necessarily identical.

The simplest approximation which could be applied consists in neglecting correlations altogether, $\WQ\equiv 0$, leading to $\dot\WP = \mcM(\WP)$. Unsurprisingly, this reduces the interacting dynamics of different partitions to its mean-field approximation.

More interesting is a less drastic approximation, also following the spirit of \emph{Sto\ss{}zahlansatz}. Consider the pragmatic assumption that the aggregated correlations expressed by $\GP$  have no bearing on practical observations if they are further in the past than a timescale $\tau^*$.

If, from the perspective of one partition, the aggregate of all other partitions forms a large reservoir with finite correlation decay time, this is well justified~\footnote{In a sense, this could be called \emph{bootstrapping} dissipation into the dynamics (in a rather gentle way)}. Equation \eref{eq:eomQ} can be thus modified such that correlations are gradually forgotten.

The simplest modification of this kind is probably linear damping, which can be expressed through the modified equation
\begin{align}
  \dot\WQ &= \Liou\WQ - \mcF(\WP,\WQ) + \Liou\WP - \mcM(\WP) %\nonumber\\
          - \WQ/\tau^*.  \label{eq:eomQdamped}
\end{align}
The goal of such an approximation need not be an effective reduced description involving only the ``relevant'' part $\WP$; $\WQ$ can be a meaningful correction. With suitable projections, and $\tau^*$ chosen properly, the sum $\WP+\WQ$ should contain enough information about correlations such that correlated systems (and echo experiments) can be accurately described.

The approximate system consisting of Equation \eref{eq:eomP} and \eref{eq:eomQdamped} can be rephrased as approximate dynamics of $W = \WP + \WQ$, with an auxiliary variable $\WP$:
\begin{align}
  \dot W &= \Liou W - (W-\WP)/\tau^*\label{eq:Wdamped}\\
  \dot \WP &= \mcF(\WP,W).\label{eq:Wpdamped}
\end{align}
For $\tau^*\to \infty$, Equation \eref{eq:Wdamped} reverts to the exact dynamics, governed by Liouville's equation, while $\tau^*\to 0$ fixes $W\equiv\WP$, i.e., the latter limit takes us back to dynamics in the mean field approximation. For all finite values of $\tau^*$, the damping term will render any correlations introduced by the interactions transient. It should be noted that Equation~\eref{eq:Wdamped} has a different meaning than Equation~\eref{eq:meandN}, since $\WP$ is no longer the instantaneous projection of $W$, but the dynamical state governed by Equation~\eref{eq:Wpdamped}, which interleaves time propagation with the projections implied by the operation $\mcF$.

Linear damping corresponds to an exponential memory window imposed on $\GP$. An interesting alternative is the use of a rectangular memory window of width $\tau^*$. The corresponding residual density matrix $\WQ$ can be obtained from a two-time function $\tilde\WQ(t,t')$ which is propagated from $\tilde\WQ(t-\tau^*,t-\tau^*)$ to $\tilde\WQ(t,t-\tau^*)$ according to
\begin{align}
\partial_t \tilde\WQ(t,t') &= \Liou\tilde\WQ(t,t') - \mcF(\WP(t),\tilde\WQ(t,t')) %\nonumber\\
+ \Liou\WP(t) - \mcM(\WP(t))\label{eq:tWQprop}
\end{align}
with boundary condition $\tilde\WQ(t',t')=0$. The approximate windowed $\WQ$ is obtained as
\begin{equation}
\WQ(t) = \tilde\WQ(t,t-\tau^*)
\end{equation}
where we set $\tilde\WQ(t,t') = \tilde\WQ(t,0)$ for $t'<0$, which is consistent with the original initial value problem \eref{eq:eomQ}.
Equation \eref{eq:eomP} for $\WP$ remains unchanged. Equation \eref{eq:tWQprop} is of the same form as the exact result \eref{eq:eomQ}, except for second time argument made necessary by the modified boundary condition.

Again, the system of equations for $(\WP,\WQ)$ can easily be rewritten as one for $(W,\WP)$, yielding
\begin{align}
\partial_t \tilde W(t,t') &= \Liou \tilde W(t,t') \label{eq:eomWti} \\
\dot\WP &= \mcF(\WP(t),\tilde W(t,t-\tau^*))\label{eq:eomP2}
\end{align}
with boundary condition $\tilde W(t',t') = \WP(t')$.

The system of equations \eref{eq:eomWti} and \eref{eq:eomP2} constitutes a further central result. Since it was constructed by discarding correlations, it can reasonably be expected to have an increase of entropy as a key feature~\footnote{A formal proof is more difficult here than for Equation \eref{eq:stochdN} or \eref{eq:meandN} since the system of equations constitutes an interleaved rather than a sequential pattern of projection and propagation.}. These equations yield $\tilde W(t,t-\tau^*)$ as an approximation to the $W(t)$ governed by the Liouville equation, displaying \emph{all} correlations which have built up in the interval $[t-\tau^*,t]$. This system of equations is therefore also applicable when protocols are available to ``rewind'' the microscopic dynamics~\cite{hahn50,schia23} by times up to $\tau^*$.

This system can obviously be given the alternate form
\begin{empheq}[box=\doublebox]{align}
\partial_t \tilde W(t,t') &= \Liou \tilde W(t,t')
\label{eq:eomWti2}\\
\partial_t \rho_j(t) &= \otr_j \Liou \tilde W(t,t-\tau^*),\quad j=1\ldots n
\label{eq:eomrho}\\
\tilde W(t',t') &= \rho_1(t')\otimes\cdots\otimes \rho_n(t')
\label{eq:bcrho}
\end{empheq}
where nonlinearity appears only in the boundary condition \eref{eq:bcrho}; the differential equations are now linear. The system \eref{eq:eomWti2}--\eref{eq:bcrho} appears to be one of the most appealing applications of the EASE principle, it will be designated as EASE-Liouville dynamics (EASE-L dynamics).

For a given $t'$, the two-time function $\tilde W(t,t')$ only needs to be propagated for $t'<t\leq t'+\tau^*$. The equivalent condition $t-\tau^* \leq t' <t$ points to a way of greatly reducing the numerical cost of two-time propagation: Since the memory truncation makes sense only in situations where the end result is insensitive to moderate changes in $\tau^*$ (for all relevant observables), one may choose a relatively coarse grid for $t'$ (not $t$), such that the grid width $\Delta\tau$ amounts to some fraction of $\tau^*$. This approach has been practiced in the context of a similar windowed Nakajima-Zwanzig formalism~\cite{stock16a,wiedm16}; it introduces $t$-dependent changes of relative size $\Delta\tau/\tau^*$ to the memory time $\tau^*$, and it leads to no more than $\tau^*/\Delta\tau$ copies of $\tilde W(t,t')$ being concurrently propagated at any time. The numerical cost of EASE-L is thus larger than for the linearly damped system \eref{eq:Wdamped} and \eref{eq:Wpdamped}, but only by a moderate factor $\tau^*/\Delta\tau$. The grid-based function $\tau^*(t)$ is bounded from below by the constant $\tau^*$ set for the continuum case.

\section{Discussion}

A description of irreversibility as an intrinsic feature of a dynamical law cannot be accomplished without some modification of the microscopic dynamics, since the latter is reversible. The EASE principle lifts the crucial element of the Boltzmann dynamics, i.e., \emph{Sto\ss{}zahlanssatz}, to a more abstract level. It states that a system which has both a dynamical and a thermodynamical description can be seen as gaining entropy by carefully backdating part of the information loss implied in thermodynamic observation to earlier times in the dynamics. This backdating is accomplished by gradually projecting away correlations during the propagation. The strategy of gradually weaving an anticipated information loss into a structure of effective dynamical laws results in dynamical states which do depend on the initial state and the fundamental microscopic laws, but the evolving distributions are not identical with conditional distributions rigidly defined by the microscopic initial value problem. This feature of EASE seems reminiscent of the ``almost-objective'' probabilities advocated by Myrvold as a general concept in statistical physics~\cite{myrvo16,myrvo21}.

The system partitions defining the correlations as well as the memory time $\tau^*$ after which correlations become irrecoverable should be adapted to the concrete system at hand. The limit $\tau^*\to\infty$ always recovers the exact microscopic description, but for $\tau^*$ too large, the EASE approach becomes impractical.
The partitions may be large in numbers or large in size, depending on the application. In the case of a finite-size thermal reservoir, the number of partitions may be as small as $n=2$~\cite{willi74}. This is of interest when a finite-size reservoir doubles as a detector~\cite{karim20}. Here EASE offers an approach where both system and reservoir observables are fully accessible; this is not the case in conventional projection or master equation approaches. The EASE principle leads to the generalized Nakajima-Zwanzig equation~\eref{eq:NZnonlin} involving only the relevant state $\WP$, an important starting point for approximations to be applied case by case. EASE-L, however provides information beyond the relevant density matrix; it additionally yields all correlations built up over the sliding window $[t-\tau^*,t]$. In extreme cases, such as many-particle localization~\cite{abani19} or many-particle systems where gravitational interactions dominate, equilibration is not expected, and the EASE description may not offer little or no advantage over a fully microscopic description.

Care must be exercised in the choice of partitions such that the projection $\mcP$ remains compatible with conservation laws, as a minimum, changes induced by $\mcP$ in any conserved quantity, in particular energy, should be slow compared to the rate at which entropy changes.

Since the full microscopic picture is recovered in the limit of large memory time, the Gibbs-Shannon-von Neumann entropy found under EASE should not be over-interpreted until an equilibrium state is reached. However, the microscopic entropy derived assuming EASE should be considered as a lower bound for any thermodynamically meaningful nonequilibrium entropy, as long as the EASE description is not made so coarse that observables are in conflict with thermodynamics~\footnote{This statement can be considered means-relative: The question of how large $\tau^*$ has to be should depend on the sophistication and accuracy of any available ``rewind protocols''~\cite{hahn50,schia23}}.

This detachment of information entropy from thermodynamic entropy is, on the other hand, a vital asset of the method: It is suitable for the description of irreversible process far from equlibrium, including processes which only settle into nonequilibrium steady states (with energy/heat sources and drains added to the picture).

The EASE principle and the EASE-L dynamics are presented in the context of quantum mechanics. The parallel classical case is easily recovered by substituting integrations over all phase space coordinates not pertaining to subsystem $j$ for the operation $\otr_j$, yielding a marginal phase-space density for this subsystem. There is, however one interesting difference between quantum and classical cases. In the quantum case, entropy augmentation should typically be observed even when starting from a pure state: Interactions will lead to entanglement between subsystems, then the entropy gain after applying $\mcP$ for the first time will be equal to the entropy of formation. In the classical case, there has to be some finite initial entropy to get entropy augmentation started.

%% furter philosophical reference on quantum-classical distinction?

The combination of linear propagation and a factorizing boundary condition in EASE-L is suggestive of a hybrid numerical approach to finite-temperature quantum many-body physics. In cases where diagonalization of the $\rho_j$ is feasible, pure states can be sampled (with the eigenvalues of the $\rho_j$ as probabilities) to yield a wave-function Monte Carlo method propagating pure states as a substitute for the propagation of $\tilde W$. This could be managed with tensor network methods, certainly with lesser difficulty than tackling the Liouville equation itself, for two reasons: On the one hand, the initial state factorized, on the other hand, the length of the propagation interval is always bounded through the constant $\tau^*$.

\section{Conclusion}
The increase of single-particle information entropy is an intrinsic feature of the Boltzmann equation, whose very structure implies decorrelation after each scattering event (\emph{Sto\ss{}zahlansatz}). Disregarding correlations immediately leads to a subadditive excess in the sum of single-particle entropies, which allows a demonstration of the H-Theorem from abstract, general features of information theory. These features can be applied to virtually any decomposition of a dynamical system into interacting subsystem, with a finite memory time imposed on inter-subsystem correlations (EASE principle). The use of a formal decorrelating projector leads to a nonlinear reduced dynamics for the projected state, similar to a Nakajima-Zwanzig equation. Alternatively, the EASE principle yields a system of linear equations, EASE-L. With current computational resources and contemporary computational methods, tensor network methods, in particular, it is likely that hybrid methods based on EASE-L will yield quantum many-particle propagation algorithms with very favorable scaling, potentially breaking the curse of dimensionality.\\[1ex]

%\subsubsection*{Acknowledgments}
\noindent\textbf{Acknowledgments}
This work was supported by the German Science Foundation (DFG) under AN336/12-1 (For2724).
The author thanks R. Balian, G. Falci, A. Nitzan, J. van Ruitenbeek and V. \v{S}pi\v{c}ka for valuable hints, useful comments and encouragement.\\[1ex]

%\subsubsection*{Data availability}
\noindent\textbf{Data availability} Not applicable.

\bibliography{ease}

\end{document}